# Pipelined Algorithms to Detect Cheating in Long-Term Grid Computations

Michael T. Goodrich

Dept. of Computer Science, Univ. of California, Irvine, CA 92697-3235 USA

**Abstract**

This paper studies pipelined algorithms for protecting distributed grid computations from cheating participants, who wish to be rewarded for tasks they receive but don't perform. We present improved cheater detection algorithms that utilize natural delays that exist in long-term grid computations. In particular, we partition the sequence of grid tasks into two interleaved sequences of task rounds, and we show how to use those rounds to devise the first general-purpose scheme that can catch all cheaters, even when cheaters collude. The main idea of this algorithm might at first seem counter-intuitive—we have the participants check each other's work. A naive implementation of this approach would, of course, be susceptible to collusion attacks, but we show that by, adapting efficient solutions to the parallel processor diagnosis problem, we can tolerate collusions of lazy cheaters, even if the number of such cheaters is a fraction of the total number of participants. We also include a simple economic analysis of cheaters in grid computations and a parameterization of the main deterrent that can be used against them—the probability of being caught.

## 1 Introduction

One of the success stories of parallel and distributed algorithms is the computational grid paradigm for solving large computational problems. In this paradigm, a *supervisor* distributes a set of independent tasks to a community of participants, who perform those tasks and send back the results. (See Figure 1.)

Examples of well-known on-going grid computations include SETI@home, which claims over 7 million participants who have collectively performed over 1.5 billion tasks aimed at finding intelligent patterns in extraterrestrial signals, and Grid.org, which claims over 3 million computers being used to solve large scientific problems related to medicine.

The participants in grid computing environments are typically volunteers rewarded with recognitions of their service. For example, SETI@home regularly posts the names of its top 1000 users. Unfortunately, even with such modest rewards, grid computations must deal with cheaters. Indeed, the director for SETI@home is quoted [15] as saying that their project spends half of their resources dealing with cheaters, who comprise roughly 1% of their users. He mentioned that some users have modified the SETI@home software to make it look like they have performed more work than they actually did. To deal with such cheaters, the SETI@home director mentioned that their system duplicates every task and sends it to two different participants for confirmation. If the two results match, they accept the computation; if the results don't match, they send the computation to a third participant for determination of which of the original two cheated. Such duplication creates waste in the system, of course, and, even with this extra cost, it still allows for cheating if participants collude. Thus, we would ideally like to find more efficient cheater deterrents that significantly discourage cheating users even if they collude. The problem of dealing with cheaters in a grid computation becomes even more serious, of course, when the rewards for participation become more tangible. Therefore, an important challenge is the efficient detection and deterrence of colluding users who wish to be rewarded for grid tasks they don't actually perform.

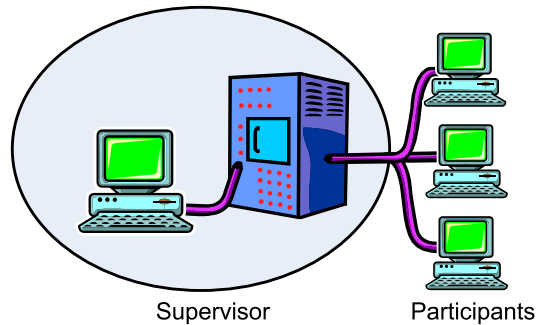

Figure 1: Illustrating the grid paradigm whereby a supervisor partitions a computation into tasks that are then farmed out to participants, who are expected to perform those tasks and submit responses back to the supervisor.

## 1.1 Previous Related Work

The research area of designing grid protocols for detecting cheating users is known as *uncheatable grid computing* [10, 11, 13, 14, 22]. Roughly speaking, previous approaches have relied either on the general-purpose approach of replicating tasks, which creates multiplicative overheads (as in the SETI@home approach), or on special-purpose ad hoc solutions. Golle and Mironov propose a special-purpose *ringer* scheme [13], which is restricted to the inversion of one-way functions. Szada, Lawson, and Owen extend their scheme somewhat [22], but their solutions still are not general purpose. Du and Goodrich [10] show how to utilize chaff-injection techniques to reduce the ability of lazy cheaters to collude on outlier-search computations. Likewise, Du *et al.* [11] propose ad hoc checking algorithms for cheater detection in grid computations. In their approach, the supervisor randomly selects and verifies for himself some samples from the task domain $D$ assigned to a participant. This approach places a significant computational burden on the supervisor in addition to his management role, since in this scheme he must redo some of each participant's tasks himself. Unfortunately, none of these solutions are general enough for all grid computations. In addition, we are not aware of any previous solutions to uncheatable grid computing that are correct in the worst case if cheating users collude.

Task duplication, which is used by SETI@home [15], is a general method for dealing with cheating users, albeit with computational overheads, since duplicating every task cuts the efficiency of the grid computation. Golle and Stubblebine [14] propose a more efficient duplication scheme, using economic disincentives for cheating. Their scheme is still not immediately applicable to most grid computations, however, as it assumes explicit payments for computations and it requires users to place a large number of unrewarded tasks on "deposit" with the supervisor prior to being paid.

There are also cryptographic protocols, such as Private Information Retrieval [8] and Probabilistically Checkable Proofs [23], that can be used to achieve uncheatable grid computing. Although such heavy machinery can provide possible theoretical uncheatable grid computing constructions, their expensive computational costs make them inappropriate choices for grid computing in practice.

Replication of tasks can be used to check arbitrary tasks, but there is also a rich body of work on ad hoc methods for checking specific kinds of computations. For example, the concept of *certification trails* for certain kinds of data structures and algorithms was introduced in [20, 21]. These are used to verify if the responses from a data structure to a sequence of operations were correct or not. Important requirements of a certification trail are that (1) generating the trail should not cause any significant overhead and (2) verifying the trail should be asymptotically faster than executing the operations on the data structure. Efficient certification trails for basic data structures, such as union-find structures and priority queues are presented in [12, 20]. An important observation made in [20] is that, in a sequence of operations, it is often more efficient to allow *latency in detection*. Namely, if an answer is incorrect, it is not necessary to detect it immediately, as long as it can be eventually detected at some later time during the verification process. Such latency has not been exploited previously for grid computations, however. Work on certification trails for graph and geometric algorithms appears in [9, 17]. Likewise, Blum and Kannan [5, 6] developed a theoretical framework for program checkers. This work, in turn, inspired a number of related papers [4, 7, 16, 18, 19]. An important distinction between this work and grid computation checking is that in the former we have a static program with possible bugs, while in the latter we have an active, possibly malicious, set of adversarial cheating participants.

## 1.2 Our Results

In this paper, we provide algorithms for cheater detection in grid computations that are based on the partition of the sequence of tasks into inter-leaved sequences of *rounds*, such that the computations in one round complete (or are timed out) before we assign the computations for the next round.

Such delays are an inherent part of grid computing environments anyway, so we feel they should be exploited to thwart cheaters.

By exploiting such grid delays, we are able to achieve the first general-purpose scheme that can catch all cheaters, even when cheaters collude. The main idea of this algorithm, which we present in Section 2, might at first seem counter-intuitive—we have the participants check each other's work. A naïve implementation of this approach would, of course, be susceptible to collusion attacks, but we show that, by adapting efficient solutions to the parallel processor diagnosis problem, we can tolerate collusions of lazy cheaters, even if the number of such cheaters is a fraction of the total number of participants. This scheme is admittedly sophisticated, however, and is practical only when the cost of even one undetected cheater is too high to tolerate. In particular, this approach requires the use of constant-degree graphs that possess a property we call $(\alpha, \beta)$-resilience, which can be constructed probabilistically, with high probability, with a linear-time Monte Carlo algorithm. In addition, while this approach requires that each task be replicated only a constant number of times (and these constants are fairly small, being only $6$ or $12$ in specific instances we give), such replication reduces the efficiency of a grid computation by the reciprocal of the replication constant.

Therefore, we also include, in Section 3, a simple economic analysis of cheaters in grid computations and a parameterization of the main deterrent that can be used against them—the probability of being caught. We show that, again by exploiting the rounds in a grid computation, we can provide simple and efficient ways of using grid delays to achieve probabilities of being caught that are arbitrarily close to $.95$, using schemes that are quite practical. This approach allows us to achieve equivalent probabilistic levels of security as previous schemes, but with lower computational overheads for the computation supervisor (the latter of which is, of course, the main goal of grid computing). These solutions could therefore improve the practical performance for the vast majority of grid computations. In particular, we provide an economic analysis of uncheatable grid computing, studying the incentives and costs for cheating in a way that avoids the use of explicit payments. Unlike the Golle and Stubblebine [14] approach, our economic model allows for implicit incentives for user participation, such as top-user recognition. In addition, our analysis includes the cost of cheating even when a user is not caught. We show that reasonably-motivated cheaters can be deterred using fewer resources than previous schemes, by making use of the rationality of cheaters and task distribution across the rounds of a grid computation.

## 2  Uncheatable Grid Computing

A grid supervisor in a long-term grid computation takes a large scientific problem, subdivides it into independent *tasks*, and iteratively sends tasks out to participants over the course of months or years. Each time an active participant completes one task, and remains available to perform tasks, he is given another task so as to continue the grid computation until it completes. (See Figure 2.)

Given this scenario and an upper bound on the time a supervisor is willing to wait for a task to be performed, we can easily partition a sequence of tasks and task responses (which have similar completion speeds) into two inter-leaved sequences of independent tasks, as shown in Figure 2. This allows us to view tasks as being performed in a series of *rounds*, so that the results of a previous round are completed, or are timed out, before the tasks for the next round are assigned. If there are multiple scales of completion speeds, e.g., when a grid includes supercomputers and laptops, then this separation into two sequences of tasks done in rounds can be done for each time scale (we assume, in this case, that there is a sufficient number of participants in each time scale for this separation to make sense). So, for the remainder of this paper, we assume there is a single time scale and that tasks can be distributed in a series of rounds at this scale. This ability of processing tasks in rounds, which is ubiquitous in grid computing but has not been previously exploited for uncheatable grid computing, allows us to assign replicated (result-checking) tasks in an adaptive fashion.

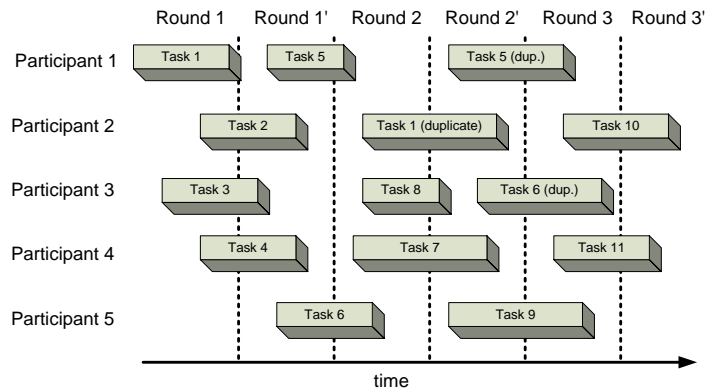

Figure 2: An example assignment of tasks to participants over time. This example illustrates how some tasks are duplicated to check previous responses. It also illustrates how time can be divided into two sequences of rounds so that tasks can start in a given round and end before the next round in that sequence begins.

There are some scenarios, such as in medical grid computations, when even a small chance of undetected cheating is unacceptable. In this section, we study *uncheatable grid computing*, which allows us to achieve perfect cheater detection, even if all cheating users collude. Our approach involves an interesting adaptation of a parallel fault diagnosis algorithms of Beigel *et al.* [2, 3] to uncheatable grid computing.

The main challenge to designing an algorithm to identify all cheating users in a grid computation is that colluding users can produce equal answers on replicated tasks, making it look like they are cooperating ("good") users. Let us, therefore, formally define the participants in a grid computation.

- *The supervisor*. This agent has a collection $\{t_1, t_2, \ldots, t_n\}$ of *tasks*, such that the supervisor would like to receive the result of a function, $g(t_i)$, performed on each task $t_i$, for $i = 1, 2, \ldots, n$. The function $g$ is assumed to be expensive, so that the supervisor desires not to compute $g(t_i)$, for any task $t_i$, himself.

- *Good participants.* Simply put, good participants do not cheat. Given a task $t_i$, a good participant computes a response $r_i = g(t_i)$ in a timely manner and returns this result to the supervisor.

- *Cheating participants.* Given a task $t_i$, a cheating participant $p$ computes a false function, $f_p(t_i)$, on $t_i$. For example, $f_p$ may be much cheaper than $g$ for $p$ to compute, which provides an incentive for $p$ to prefer to perform $f_p$. Alternatively, $f_p$ may hide valuable information that $g$ would otherwise report back to the supervisor. In any case, the only assumption we make about these functions is that $f_p(t_i) \ne g(t_i)$, for otherwise there is no actual occurrence of cheating on the task $t_i$.

Note that cheating users may sometimes produce identical results. The goal of uncheatable grid computing, then, will be to detect each instance of cheating. That is, we want to detect each time there is a response $r_i$ for a task $t_i$ such that $r_i \ne g(t_i)$, even if cheating participants try to "cover each other's tracks."

We assume in this section that there is no *traitor* $p$ in the group of participants, who would correctly compute $g(t_i)$ for a task $t_i$ initially assigned to $p$ but who also would replicate the response of a false function $f_q(t_i)$ on $t_i$ if the task $t_i$ were initially assigned to a cheating participant $q$. That is, there are no users who produce correct responses themselves but are willing to corroborate incorrect responses from cheaters[1]. In fact, it is easy to show that if there are $k > 0$ traitors, out of $n$ participants in a grid computation, then a testing algorithm that is 100% correct in the worst case requires at least $\Omega(k)$ rounds, and even a probabilistic algorithm that is correct with high probability requires at least $\Omega(\log_{n/k} n)$ rounds.

We begin our checking protocol by having each user $i$ commit to a response $r_i$ to a task $t_i$ in a preliminary round. Then, in subsequent rounds, the supervisor mixes in testing tasks, which are replications of $t_i$ sent to other participants. Recall that we assume that a false response $r_i$ from a cheating user will not match the correct response $r'_i = g(t_i)$ when the task $t_i$ is performed by an honest user. But if the task $t_i$ is assigned to another cheating user, we allow for these users to collude so that the second respond matches the first (incorrect) one. That is, two colluding cheating users $p$ and $q$ could arrange for $f_p(t_i) = f_q(t_i) = r_i$. Since the only means the supervisor has to detect a false response is when two responses on the same task differ, he can not immediately detect a collusion such as this.

## 2.1 A Reduction to Parallel Fault Diagnosis

This response semantic provides a relationship of this problem to the parallel fault diagnosis problem. In this related problem, we are given a set of $n$ processors, each of which is either "good" or "bad." In a single round, a processor can test another processor or be tested itself by another processor. If a good processor tests another processor, it correctly returns to a central supervisor whether the tested processor is good or bad. On the other hand, if a bad processor tests another processor, it returns an arbitrary (or even deliberately false) identification of the other processor as being good or bad. The fault diagnosis problem is to determine all the good and bad processors using a small number of parallel testing rounds, assuming some upper bound on the number of bad processors (the problem cannot be solved if there are more than half as many bad processors as good).

---

[1] In our more practical algorithm given later, in Section 3, we can allow for traitors as well as cheating users and still have efficient performance, albeit at the expense of uncheatability.

We can adapt a parallel fault diagnosis solution to become a solution to the problem of checking the results from a grid computation performed for $n$ tasks. For a *test* in our case corresponds to a supervisor distributing the task $t_i$ performed by a participant $p$, resulting in the answer $r_i$, to another participant $q$ (with $q \neq p$) and receiving a result $r'_i$ from $q$. If both responses are good, the results will match, i.e., we will have $r_i = r'_i$. If one of them is good and the other is false, the results will not match, i.e., we will have $r_i \neq r'_i$. But if both participants are cheaters, however, the results can match, even if they both cheat, just by having the two participants collude, so we may have $r_i = r'_i$ in this case.

To use a parallel fault tolerance algorithm in this context, then, we can proceed as follows. After a preprocessing round, which distributes the $n$ tasks to participants and receives their results, we can then simulate the parallel fault tolerance algorithm by replacing each test of a processor $i$ by a processor $j$ with a replication of the task sent to participant $p$ with a copy sent to a participant $q$ and then have the supervisor simply test the equality of the results for each such pair and use the fault-detection protocol to partition the results into sets.

For example, the 10-round algorithm of Beigel *et al.* [2] immediately translates into a 10-round checking algorithm (after the preprocessing round that commits the results of the tasks). As we show below, however, we can improve this performance by being more careful in how we distribute duplicated tasks.

## 2.2 Improved Protocols for Uncheatable Grid Computing

There are some improvements that we can make to the above approach, however, for uncheatable grid computing. First, in any round, we can allow a participant $q$ to test another task $p$ even if $q$'s task is also being tested in that round (such tests are not allowed in the parallel fault diagnosis problem). This observation lets us immediately reduce the number of rounds in our simulation to 9, since the first round in the algorithm of Beigel *et al.* involves the symmetric testing of $n/2$ pairs of processors (which requires two rounds in their algorithm but only one in our simulation). Even so, this simulation algorithm is probabilistic and only guaranteed to succeed with high probability if $n$ is very large. Thus, we would like to reduce further the number of rounds and make this algorithm work with high probability for reasonable values of $n$. To achieve these goals, let us make a simplifying assumption, which is well-motivated for grid computing but not for parallel fault diagnosis: namely, let us assume that the number of cheating participants is much less than $n/2$. The motivation for this assumption is that the pipelined nature of grid computing allows the supervisor to prune away cheating participants as soon as they are discovered (by our testing algorithm); hence, it is unlikely for large numbers of cheaters to be in the grid.

We use constant-degree directed graphs in our pipelined algorithms, where we define the *degree* of a directed graph to be the maximum in-degree or out-degree of its vertices. Say that an $n$-vertex directed graph $G$ is $(\alpha, \beta)$-*resilient* if any subgraph of size at least $\alpha n$ has an induced strongly-connected component of size at least $\beta n$. We show later in this section there is an $n$-vertex, degree-4 directed graph $H_4$ that is $(15/16, 7/16)$-resilient.

Let us illustrate how to use the existence of such a graph, $H_4$, in a 3-round testing algorithm, which will catch all cheaters so long as at most 5% of our participants are colluding cheaters (which is certainly achievable for a supervisor who is actively removing cheaters). In particular, our algorithm is based on the existence of a directed graph, $H_4$, such that no matter how an adversary might distribute "bad" vertices in $H_4$, there is a large connected component consisting exclusively of good vertices. Our protocol for catching all forged answers using $H_4$ in a set of $n$ tasks is as follows.

1. View each participant (and his task) as a vertex, and divide the set of these $n$ vertices into $n/4$ directed cycles of $4$ vertices each. In one round have each vertex test the next one in the cycle, using task replication. Discard for now all the vertices in any cycle with a detected bad vertex (there can be at most $4n/20 = n/5$ such discarded vertices). Let $N \geq (n - n/5)/4 = n/5$ denote the number of remaining cycles, which we view as super-vertices, since all the remaining cycles each consist entirely of good vertices or bad vertices.

2. There can be at most $(n/20)/4 = n/80$ such all-bad cycles, that is, at most $5N/80 = N/16$ bad super-vertices. Let us now assume we have a degree-4, $N$-vertex directed graph, $H_4$, that is $(15/16, 7/16)$-resilient. Apply the tests dictated by the edges of $H_4$ in one round, viewing the $N$ super-vertices as the vertices in $H_4$. (This testing is done by having the four participants in the origin super vertex repeat the tasks for the four respective participants in the destination super vertex.) Note that, by the definition of $H_4$ and the fact the number of bad super-vertices is small enough, there will be a strongly connected component of $(7/16)N$ super-vertices in $H_4$ (which must necessarily be all good, since there are not so many bad vertices). That is, we will have at most $(9/16)N \leq (9/16)n/4 = (9/64)n$ super-vertices whose classification may be in doubt.

3. Note that at this point there are at least $(7/16)4N \geq (7/4)n/5 = (7/20)n$ identified good vertices. Divide these vertices into two groups: one group of size $n/5 = (4/20)n$ vertices, which are used in one round to test the previously-discarded vertices, and another group of $(3/20)n = (9/60)n$ vertices, each of which can test a single super-vertex in one round (since $9/60 > 9/64$).

Thus, we have the following.

**Theorem 1** *Given a method for finding an $N$-vertex, degree-$4$ directed graph $H_4$ that is $(15/16, 7/16)$-resilient, for $N \geq n/5$, one can identify in three rounds all false replies among $n$ tasks in a grid computation, so long as there are at most 5% bad participants.*

Note, as well, that the replication factor of any task in the above algorithm is six.

## 2.3 Finding Resilient Graphs

In completing the details of our grid computing testing algorithm above, we obviously need to have a method for finding $(15/16, 7/16)$-resilient directed graphs of degree-$4$. To help us perform this important step, we utilize the following lemma from Beigel *et al.*:

**Lemma 2 ([2])** *Let $G = (V, E)$ be a directed graph on $n$ vertices. Let $0 < \lambda, \gamma < 1$. Suppose, for every pair of subsets $A$ and $B$ of $V$ such that $A \cap B = \emptyset$, $|A| + |B| = \lambda n$, and $|A|, |B| \leq \frac{1+\gamma}{2} \lambda n$, that there are edges in $E$ directed from $A$ to $B$ and $B$ to $A$. Then $G$ induces a strongly connected component of size $\gamma \lambda n$ on any subgraph with $\lambda n$ vertices.*

This lemma may at first seem obscure, but it is useful for proving the following theorem, which extends a theorem from an earlier work of Beigel *et al.* [3].

**Theorem 3** *Let $V$ be a set of $n$ vertices, and let $0 < \gamma, \lambda < 1$. Let $H_d = (V, E)$ be a directed graph defined by the union of $d$ independent randomly-chosen[2] Hamiltonian cycles on $V$ (with all such cycles equally likely). Then, for all subsets $W$ of $V$ of $\lambda n$ vertices, $H_d$ induces at least one strongly connected component on $W$ of size greater than $\gamma \lambda n$, with probability at least*

$$1 - e^{n[(1+\lambda)\ln 2 + d(\alpha \ln \alpha + \beta \ln \beta - (1-\lambda)\ln(1-\lambda))] + O(1)},$$

*where $\alpha = 1 - \frac{1-\gamma}{2}\lambda$ and $\beta = 1 - \frac{1+\gamma}{2}\lambda$.*

**Proof**: The proof is an adaptation and correction of a proof of a weaker theorem from Beigel *et al.* [3]. By Lemma 2, it is sufficient to show that with the exponentially small probability mentioned in Theorem 3, there is a subset $W$ of $V$ of size $\lambda n$ that has a partition $(A, B)$, with $|A|, |B| \leq \frac{1+\gamma}{2}\lambda n$, such that there is no edge from $A$ or $B$ or no edge from $B$ to $A$. Let us consider first the probability that there is no edge from $A$ to $B$ (as the other case is identical). Beigel *et al.* [3] show that, for a single randomly-chosen Hamiltonian cycle $H$ on $V$ (and two disjoint subsets $A$ and $B$ of $V$), the probability that there is no edge from $A$ to $B$ is

$$\frac{(n-|A|)!(n-|B|)!}{n!(n-|A|-|B|)!}.$$

Thus, the probability that there is no edge from $A$ to $B$ or no edge from $B$ to $A$ in $H_d$ is at most

$$2\left(\frac{(n-|A|)!(n-|B|)!}{n!(n-|A|-|B|)!}\right)^d.$$

There are at most $2^n$ choices for $W$ and at most $2^{\lambda n}$ possible ways of partitioning $W$ into subsets $A$ and $B$ (actually, there are fewer, but these bounds will suffice for our purposes). Thus, the probability that there is a subset $W$ of $\lambda n$ vertices that has a partition $(A, B)$, with $|A|, |B| \leq \frac{1+\gamma}{2} \lambda n$, such that there is no edge from $A$ or $B$ or no edge from $B$ to $A$ is at most

$$2^{(1+\lambda)n+1} \left( \frac{(n-|A|)!(n-|B|)!}{n!(n-|A|-|B|)!} \right)^d.$$

---
[2]That is, $H_d$ is defined by the union of cycles determined by $d$ random permutations of the $n$ vertices in $V$, so $H_d$ is, by definition, a simple directed graph.

This is maximized when $n - |A| = \alpha n$ and $n - |B| = \beta n$. Applying Stirling's formula, we can bound this probability by

$$e^{n[(1+\lambda)\ln 2 + d(\alpha \ln \alpha + \beta \ln \beta - (1-\lambda)\ln(1-\lambda))] + O(1)}.$$

∎

Applying this to our testing problem, we need to set the parameters so that we are guaranteed to have at least $\lambda n$ good testers, for using a directed graph $H_d$ for sufficiently large $d$ will guarantee with high probability that the subgraph of good testers will induce a strongly connected component of size at least $\gamma \lambda n$. For example, we have the following:

**Corollary 4** *If $\gamma = 7/15$, $\lambda = 15/16$, $d = 4$, and $n \geq 20$, then any subgraph of $(15/16)n$ vertices of a random directed graph $H_d$, defined as above, induces a strongly connected component of size $(7/16)n$ with probability at least*

$$1 - e^{-n/4}.$$

Thus, we have the following:

**Theorem 5** *There exists a linear-time Monte Carlo algorithm that constructs, for any $n \geq 20$, an $n$-vertex, degree-4 directed graph, $H_4$, that is $(15/16, 7/16)$-resilient with high probability.*

## 2.4 Coping with Larger Colluding Groups

If more than 5% of the participants are colluding, the above algorithm is not guaranteed to succeed. Nevertheless, we can adapt our solution to tolerate such large coalitions, albeit at a greater expense of replicating tasks. Let us, for example, consider a case where we could have as many as 10% of the participants being colluding users.

For example, we have the following additional corollaries to Theorem 3.

**Corollary 6** *If $n \geq 20$, $\gamma = 1/2$, $\lambda = 7/8$, and $d = 8$, then any subset of $(7/8)n$ vertices induces a strongly connected subgraph of $H_d$, defined as above, of size $(7/16)n$ with probability at least*

$$1 - e^{-n}.$$

**Corollary 7** *There exists a linear-time Monte Carlo algorithm that constructs, for any $n \geq 20$, an $n$-vertex, degree-8 directed graph, $H_8$, that is $(7/8, 7/16)$-resilient with high probability.*

If we can safely assume that the number of cheating participants is at most 10% of the total (which is ten times higher than the SETI@home experience), then we can use this corollary to design the following five-round testing strategy:

1. Pair up participants and have them test each other. Discard for now any pairs that have an identified bad test (for one of them must be bad). The remaining pairs must each consist of two good participants or two bad ones.

2. Pair up pairs of participants from the first round and have them test each other with one test per participant. Discard for now any groups that have an identified bad test (for two of the four must be bad).

3. Pair up groups of participants from the previous round and have them test each other. Discard for now any super-groups that have an identified bad test (since four of the eight must be bad). Let $N$ be the number of super-groups.

4. Note that there can be at most $n/5$ discarded nodes; hence, $N \geq (n - n/5)/8 = n/10$. Moreover, since each super-group has all good nodes or all bad nodes, the number of all bad super-groups is at most $n/80 \leq N/8$. Apply the $H_8$ strategy to the super-groups, where $H_8$ is constructed as in Corollary 7. This results in a strongly connected component (which must be all good nodes) of size at least $(7/16)N$, which is at least $(7/16)(n/10)8 = (7/20)n$. Moreover, there are at most $(9/16)N$ unresolved super-groups, which is at most $(9/128)n$ super-groups.

5. Split the $(7/20)n$ proven good nodes into two groups: one group of size $n/5 = (4/20)n$, which is sufficient to have each test a discarded node, and another group of size $(3/20)n$, which is sufficient to have each test a representative member of an unresolved super-group (since $3/20 = 9/60 > 9/128$).

Thus, we have the following:

**Theorem 8** *Given a method for finding an $N$-vertex, degree-8 directed graph $H_8$ that is $(7/8, 7/16)$-resilient, for $N \geq n/10$, one can identify in five rounds all false replies among $n$ tasks in a grid computation, so long as there are at most 10% bad participants.*

Or, put another way, we have the following:

**Theorem 9** *There is an algorithm for identifying in five rounds all false replies among $n$ tasks in a grid computation, so long as there are at most 10% bad participants. Moreover, this algorithm can be defined by a linear-time Monte Carlo process that is correct with high probability.*

Note that the replication factor of any task in the above algorithm is 12.

# 3 The Economics of Cheating in Grid Computations

In some cases, a grid supervisor can tolerate a moderate amount of cheating, provided he or she can be confident that cheaters will be caught and removed from the group of grid participants. This perspective allows us to economically model the problem of eliminating cheaters in a grid computation and use simpler testing schemes to catch cheaters in a small number of rounds.

Following the example of Becker [1] of applying economic analysis to human behavior, we can derive a simple economic model for a potential cheater. We assume that users are *rational*, wishing to maximize their expected utility return for participating in a grid computation. Such users will cheat on a computational task only if the expected return for cheating is at least the expected return for cooperating (risk-preferring users will cheat even when these values are equal). Viewed economically, the goal of a supervisor, then, is to deter cheating by setting (and advertising) the mechanism of the grid computation so that rational users will have little, if any, incentive to cheat. Important parameters that reflect on this task include the following:

- $P$, the probability of a cheater being caught cheating on a task.

- $B$, the net economic benefit for cooperatively performing a task. This benefit can include positive factors, such as the personal satisfaction derived from helping in the project goal and explicit rewards, such as website recognition. It can also include negative factors, such as the computational cost for performing a task. The "currency" with which we measure $B$ (and the parameters that follow) is not critical; what is critical is that we have a way of measuring benefits (and costs) accurately. In any case, we may safely assume $B > 0$, for otherwise, no user will ever perform any task.

- $U$, the net economic utility for cheating on a given task and not getting caught. This utility can include the personal satisfaction derived from hurting the project goal as well as explicit rewards, such as website recognition for (false) task completion. We also assume that $U$ is net any costs of cheating irrespective of whether the user is caught. Furthermore, we assume that if a cheater is caught, then his task is reassigned to another user; hence, the positive components of $U$ are realized only if a cheater is not caught.

- $C$, the cost of cheating and being caught. This cost includes any explicit fines or implicit penalties, such as being removed from the group of participants. It also includes the costs for cheating irrespective of whether the user is caught.

Given these parameters, we can easily characterize the condition that causes a rational user to avoid cheating, which occurs when her expected return for cooperating is greater than her expected return for cheating, that is, when

$$B \;>\; U(1-P) - CP.$$

Note that, since $U$ is a net utility, and $C$ is an all-inclusive cost, this equation fully captures the cost of cheating irrespective of whether the user is caught.

We can express the optimization condition for the supervisor as that of ensuring
$$P > \frac{U - B}{U + C}.$$

Notice that if $B \geq U$, the rational user will never cheat, even if he can get away with it. This is the same economic disincentive, for example, for photocopying a book that is cheaper to buy than to copy. If a supervisor can set up his grid computation so that this condition holds, then he need not concern himself with the detection of rational cheaters or the penalties for cheating. Unfortunately, given the experience of SETI@home, it is unrealistic to assume that $B \geq U$, so let us suppose the more typical scenario occurs—namely, that $B < U$. Thus, to create a disincentive for cheating it is sufficient for
$$P > \frac{1}{1 + C/U}.$$

One conclusion we can immediately draw from this is that, if $C \geq U$, then it is sufficient for $P = 1/2$ in order for us to deter cheating. For example, if the utility for cheating is based only on a user seeing his name appear on a list of top participants, then removing a cheating user from the group of participants guarantees $C \geq U$ for him. Of course, since most sets of participants in grid computations are volunteers and cheating is not a criminal act, the removal of a rogue participant is probably the main component of $C$. Thus, we might not necessarily be able to assume that $C \geq U$, but even in such cases the supervisor can probably find a lower bound for the ratio $C/U$, which will give a non-trivial effective bound for $P$ even in such cases. Therefore, the main focus of efficient cheater deterrence in grid computations can in most cases be reduced to the problem of setting a good value for $P$.

## 3.1 Applying Economics to the Supervisor in a Computational Grid

So far we have addressed the costs and utilities of potential cheaters. Let us now introduce some parameters for the supervisor:

- $L$, the loss incurred if a task is not done.
- $S$, the cost to ship, monitor, and receive a task.
- $r(P)$, the replication factor needed in order to achieve a given value of $P$.

The goal of the supervisor should be to balance the expected loss for a task not being done with the cost for replicating tasks so as to achieve a given value of $P$. That is, the supervisor should strive to achieve
$$L \cdot \Pr\left(U \geq \frac{B + CP}{1 - P}\right) = (1 + r(P))S,$$
where we view $B$ and $C$ as being fixed values. Thus, the problem of determining an effective value for $r(P)$ can be reduced to that of estimating the probability distribution on $U$, which should be a reasonable calculation for the supervisor based on polling and prior experience. Thus, the remainder of this paper studies various efficient ways of achieving values for $P$ with small replication factors.

# 4  Efficient Task Replication

As mentioned in the introduction, to test whether a task is performed correctly, a common way that some grid supervisors deal with cheating is to send the same task to multiple users. In such applications the supervisor typically accepts a result as correct if the results match.

## 4.1  The Drawback with Simple Replication

Unfortunately, as we explored in the section (2) on uncheatable grid computing, if users collude, this assumption may be false, since colluding users assigned the same task can both return the same (wrong) result. For example, if 5% of participants are colluding cheaters or traitors, all tasks are replicated once, and replicated tasks are sent out in the same round as the tasks they duplicate, then an expected 0.25% of all tasks could have forged results (for the probability of being caught cheating on these tasks is $0$). Moreover, this forgery occurs even with the full duplication of all tasks. Golle and Stubblebine [14] propose a more efficient scheme, where tasks are duplicated according to an exponential distribution, but their solution still allows colluders to tell with high probability whether they will be caught cheating or not.

We propose here a simple remedy—if a task is to be duplicated, then a supervisor should send out the "sanity-check" duplicate only after he has received the result from the original, using the same partitioning of tasks into rounds used in the algorithm of Section 2. This simple solution forces a cheater to commit to whether she will cheat prior to knowing if a duplicate task will be assigned to a co-conspirator. In the case of 1% colluding participants (i.e., cheaters and traitors), with full duplication of all tasks, this approach implies a probability of being caught of $P = .99$. Indeed, if the supervisor has an upper bound, $G$, on the fraction of non-colluding participants, then we can set a replication probability of $P/G$ and achieve a probability $P$ of catching cheaters, provided $P \leq G$.

As mentioned above, under the reasonable assumption[3] that $C \geq U$, we can deter cheating by setting $P = 1/2$. Using a reasonable factor of $G = .95$, which would imply a level of cheating that is potentially five times greater than that experienced by SETI@home, setting $P = 1/2$ implies we can deter cheating with a replication probability of only 52.6%. And this is on a per-task basis. If we make the natural assumption that a user's cheating utility remains constant during all rounds of a grid computation, then repeat cheating is likely, and each cheating attempt would be caught with equal probability, $P$. Thus, with $P = 1/2$, the supervisor is likely to catch (and remove) a cheating user with 99.9% probability after ten cheating attempts.

Having said that, we note that there still is a risk to a "Sybil" attack, whereby a user creates many pseudonyms and uses each to cheat (indeed, it might be a Sybil attack that allows "different" users to be colluding on their task responses). To help mitigate such attacks, a supervisor can use a higher replication factor for tasks given to new users, with that replication factor converging to $P/G$ as the user commits more and more correct results. For example, the supervisor could use a

---

[3]If the supervisor can only provide a non-trivial lower bound for $C/U$, then we should replace the "$1/2$" in this discussion with the value for $P$ implied by this lower bound.

replication probability of
$$\max\left\{1 - \frac{t}{20}, \frac{P}{G}\right\},$$
where $t$ is the number of completed tasks. Such a replication probability allows one to gradually go from a cheating detection rate of $G$ to one equal to $P$. In addition, note that this is different than requiring users to deposit unrewarded tasks, since in the supervisor can reward task performance immediately. Even so, there still may be some computations with a high loss cost or high cheating utility, for which a cheating detection rate of $G$ is still not high enough. For such scenarios, we may have to fall back to our worst-case solution, given in Section 2, which achieves $P = 1$ under the reasonable assumption that $G \geq .95$.

## 5 Conclusion and Future Directions

We have shown how to use pipelining to allow participants in a grid computation to check each other's work, even in the presence of arbitrary collusion among lazy cheaters (provided the number of cheaters is not too high). Note, in addition, that since the failure probability of Corollary 4 and, hence, Theorem 5, is exponentially small, our approach for uncheatable grid computing yields efficient Monte Carlo algorithms that are correct with high probability, which, in turn, implies the existence of efficient deterministic testing schemes. Nevertheless, it is not clear how to deterministically test if a graph is $(\alpha, \beta)$-resilient in polynomial time for constants $\alpha$ and $\beta$; hence, a different approach would have to be used if one desires a deterministic polynomial-time testing scheme.

For future work, it would be interesting to design learning models and probabilistic weights to score participants on their likelihood of cheating and then tailor uncheatable grid computing schemes to these scores.

### Acknowledgment

We thank Wenliang (Kevin) Du for several helpful discussions on the topic of this paper. This research was supported by NSF Grants CCR-0225642, CCR-0311720, CCR-0312760, OCI-0724806, and IIS-0713046.